\documentclass[aps,pre,onecolumn,
superscriptaddress,groupedaddress,showpacs]{revtex4}
\usepackage{graphicx,epsf}
\usepackage{psfrag}
\begin{document}
\title{Conformational transitions in random heteropolymer models}
\author{Viktoria Blavatska}
\email[]{E-mail: Viktoria.Blavatska@itp.uni-leipzig.de; viktoria@icmp.lviv.ua}
\affiliation{Institut f\"ur Theoretische Physik and Centre for Theoretical Sciences (NTZ),\\ Universit\"at Leipzig, Postfach 100920,
D-04009 Leipzig, Germany}
\affiliation{Institute for Condensed
Matter Physics of the National Academy of Sciences of Ukraine,\\
79011 Lviv, Ukraine}
\author{Wolfhard Janke}
\email[]{E-mail: Wolfhard.Janke@itp.uni-leipzig.de}
\affiliation{Institut f\"ur Theoretische Physik and Centre for Theoretical Sciences (NTZ),\\ Universit\"at Leipzig, Postfach 100920,
D-04009 Leipzig, Germany}
 
\begin{abstract}
We study the  conformational properties of heteropolymers containing two types of monomers $A$ and $B$,  modeled 
as self-avoiding random walks on a regular lattice. Such a model can describe in particular the sequences of hydrophobic 
and hydrophilic residues in proteins  (K.F. Lau and K.A. Dill, Macromolecules {\bf 22}, 3986 (1989)) and 
polyampholytes with oppositely charged groups (Y. Kantor and M. Kardar, Europhys. Lett.~{\bf 28}, 169 (1994)).
Treating the sequences of the two types of monomers as quenched random variables, we provide a systematic analysis 
of possible generalizations of this model. To this end we apply the pruned-enriched 
Rosenbluth chain-growth algorithm (PERM), which allows us to  obtain the phase diagrams of extended and compact states coexistence as function of both the temperature and 
fraction of $A$ and $B$ monomers along the heteropolymer chain. 
\end{abstract}


\maketitle%

\section{Introduction}

Many polymers encountered in chemical and biological physics can be considered as long flexible chains of linear architecture.  In the good solution regime, 
where the steric hard-core repulsions between monomers are dominant, a typical polymer chain usually attains a coil-like conformation. In this regime, the size measures of a typical $N$-monomer 
chain such as the mean-square end-to-end distance $\langle R_e^2 \rangle$ or the mean-square radius of gyration $\langle R_g^2 \rangle$ obey scaling laws  \cite{deGennes79,desCloizeaux,Grosberg94}:
$$
\langle R_e^2 \rangle  \sim \langle R_g^2 \rangle  \sim  N^{2\nu_{{\rm coil}}}, \label {RR}
$$
with a universal exponent $\nu_{{\rm coil}}=0.587597(7)$  in three dimensions \cite{Clisby00}.
At temperatures below the so-called $\Theta$-point the polymer radius shrinks, which 
causes the collapse transition into the compact globule-like conformation regime. Exactly at the $\Theta$-temperature, the macromolecules in three dimensions behave like an ideal Gaussian chain 
(up to logarithmic corrections) with size exponent $\nu_{\Theta}=1/2$.

The conformational properties of long flexible polymer chains in the vicinity of the $\Theta$-point are perfectly captured within the lattice model of self-attracting self-avoiding walks (SASAWs) \cite{Privman86,Lam90,Foster92,Barkema98,Szleifer92,Grassberger95,Grassberger97,Vogel07}.    
Here, each monomer is treated as a point particle on a regular lattice, and interactions are restricted to an attractive coupling $\varepsilon_{ij}=-1$ 
between non-bonded monomers  $i$ and $j$ occupying nearest-neighbor sites.

A subject of great interest is the study of conformational transitions of  heteropolymer chains, containing monomers of essentially different chemical nature.
Typical examples are proteins, consisting of sequences of amino acid residues, connected by peptide bonds. 
One of the ways to analyze this problem is via a generalization of the SASAW model by assuming interaction energies $\varepsilon_{ij}$ of all the 
couples of monomers along the chain to be independent identically distributed quenched random variables (so-called random-bond model \cite{Garel88,Shakhnovich89,Archontis94}). 
This in particular allows an analytic description of the thermodynamics of heteropolymers within the random-energy model  
developed by Derrida \cite{Derrida81} for the theory of spin glasses.
In Ref. \cite{Gutin93}, a modification of this model was proposed, where the energies of different 
states are random variables but may take only discrete values. 
However, more realistic is the case where the  interaction energies are determined by the underlying monomer sequence.  
For simplicity, we will focus on copolymers, when there 
are only two distinct types of monomers: $N_A$ monomers of type $A$ and correspondingly $N_B=N-N_A$ monomers of type $B$. 
One distinguishes between the ``sequence space'' (the set of all possible sequences of $A$ and $B$ residues along the chain) and the ``conformational space"
(the set of all possible internal conformations of a chain due to different bond orientation, taken at fixed sequence of $A$ and $B$). It is convenient to introduce the parameter 
$c\equiv N_A/N$ (fraction of $A$ monomers), so that $c=1$ $(0)$ corresponds to the limiting case of homogeneous chains with all monomers of type $A$ ($B$). 
In the following we shall focus on the following cases:
\begin{eqnarray}
&&{\rm HP\, model}: \varepsilon_{AA}=-1, \varepsilon_{BB}=\varepsilon_{AB}=0, \label{mod1} \\
&&{\rm Symmetrized\,HP\, (SHP)\, model}:\varepsilon_{AA}=\varepsilon_{BB}=-1, \varepsilon_{AB}=1, \label{mod2}  \\
&&{\rm Polyampholyte\, (PA)}: \varepsilon_{AA}=\varepsilon_{BB}=1, \varepsilon_{AB}=-1, \label{mod3}\\
&&{\rm Polyectrolyte\,(PE)}: \varepsilon_{AA}=1, \varepsilon_{BB}=\varepsilon_{AB}=0, \label{mod4} \\
&&{\rm Antisymmetrized\, model\,(AS)}: \varepsilon_{AA}=-1, \varepsilon_{BB}=1, \varepsilon_{AB}=0. \label{mod5}
 \end{eqnarray}
 Case  (\ref{mod1}) refers to the (minimal) HP model, proposed in \cite{Lau89} to describe  protein folding. 
In general, the constituents (monomers) of macromolecules in an aqueous environment 
can be characterized as hydrophilic (polar) or hydrophobic, depending on their chemical structure. Hydrophilic residues ($B$) tend to form hydrogen bonds 
with surrounding water molecules, whereas the hydrophobic monomers ($A$) effectively attract each other and tend to form a dense hydrophobic core. 
Despite is simplicity, this model (and its off-lattice ``AB'' model variant) catches the main thermodynamic features of real proteins, allowing to explore  the full conformational and sequence spaces of macromolecules by 
powerful numerical methods, and thus still attracts attention of researchers \cite{Shakhnovich90,Lattman94,Stillinger95,Yue95,Irback97,Irback97a,Hsu03,Bachmann03,Mezard04,Bachmann05,Bachmann07,Wust08,Swetnam09,Wust12,Wust13,Narasimhan12}. 
The thermodynamics and ground-state search  for various HP sequences were analyzed specifically, 
either representing  real proteins (e.g., a 103-mer for Cytochrome \cite{Lattman94})
or just some model cases (like Fibonacci sequences \cite{Stillinger95,Hsu03}).
The largest sequence explored systematically so far is a 136mer \cite{Yue95}.
In general, the ability of the chain to fold into a compact state was found to be dependent both on the composition (the fraction of residues of each type) 
and on the sequence of residues.

The related case (\ref{mod2}) is the standard model for copolymers with monomers that have a tendency to segregate \cite{Sfatos97}. 
In Ref.~\cite{Mezard04}, the thermodynamic properties of copolymers (1), (2) with correlated sequences were analyzed analytically, using an approach based on the cavity method as used in various frustrated systems. 
It was shown, that heteropolymers posses different dense phases depending upon the temperature, the nature of the monomer interactions, and the sequence correlation. 
 
The situation (\ref{mod3}), when the like monomers repel and the opposite ones attract each other, may refer to the randomly charged model of polyampholytes \cite{Tanford61,Dobrynin04}: heteropolymers, comprising 
both positively and negatively charged monomers. Typical examples of polyampholytes are synthetic copolymers bearing acidic and basic repeat groups.  
In general, the long-range nature of the electrostatic Coulomb interaction between charged groups produces crucial effects on macromolecule properties  
and leads to  an extremely  rich conformational behavior, strongly dependent both on their overall 
charge $Q$ and the quality of solvent (and thus on temperature $T$)  \cite{Kantor92,Gutin94,Kantor95,Dobrynin96,Barbosa96,Tanaka97,Everaers97,Lyubin99,Yamakov00}. 
 If positive and negative charges are almost balanced ($Q$ is small), the attractive 
Coulomb interaction dominates, and the polymer collapses into a globular (sphere-like) state at low temperatures. 
If the energy of the electrostatic interactions  prevails the globular surface energy,
 which happens when the excess charge $Q$ approaches  $Q_c\sim N^{1/2}$, the molecules split into extended necklace-like conformations of small connected globules. 
In the high-temperature regime but small $Q$ (predominant electrostatic attraction), the PA  forms  globule-like structures, whereas at the critical value of the 
excess charge $Q_c\sim N^{1/2}$ the charged polymer   
attains the shape of a neutral polymer coil.   
Polyampholytes with considerable disbalance between positive and negative charges ($Q>Q_c$) and thus 
predominant Coulomb repulsion  are expected to attain a 
stretched rod-like configuration.  
The simplified polyampholyte model with strongly screened (short-range) Coulomb interactions  in the form (\ref{mod3}) 
have been studied previously in Ref.~\cite{Kantor94}. Here, for each $c$ value  at fixed total length of the chain, 
the various possible sequences have been constructed, and
quenched sequence averaging for thermodynamic observables was performed. In such a way, the phase diagram of extended and compact state coexistence has been constructed
as function of both the temperature and the fraction of residues of each type.

The situation (\ref{mod4}) may describe polyelectrolytes \cite{Tanford61,Dobrynin04}:  heteropolymers, comprising some fraction of charged monomers of the same sign 
with strongly screened Coulomb interactions. 
A transition into the compact state can not be reached for these systems. 

The case (\ref{mod5}) could be viewed as an additional generalization of above models, comprising an interplay between different kinds of short-range interactions.

In the present work, we aim to extend the picture, developed in Ref.~\cite{Kantor94} for the polyampholyte model (\ref{mod3})  to the whole 
range of cases (\ref{mod1})-(\ref{mod5}); namely, to treat different sequences of $A$ and $B$ monomers at given fixed fraction $c$ (ranging from 0 to 1)
as different realizations of quenched disorder, and to obtain observables of interest after performing double conformation and sequence averaging. Thus,
we will obtain a systematic picture of conformational transitions from extended into compact state in models (\ref{mod1})-(\ref{mod5})
in the form of phase diagrams as function of the fraction $c$ and temperature $T$. 
In particular, this allows us to compare the properties of HP-like models and models of charged polymers
chains with strongly-screened Coulomb interaction directly within the same approach. 
 
The layout of the rest of the paper is as follows: 
in the next Section II we briefly describe the pruned-enriched Rosenbluth method, applied in our study. The results obtained are given in 
Section III. We end up by giving conclusions and an outlook in Section IV.

\section{Method}

To study the conformational properties of models (\ref{mod1})-(\ref{mod5}), we apply the pruned-enriched Rosenbluth method (PERM) \cite{Grassberger97}. The chain grows step by
step, i.e., the $n$th monomer is placed at a randomly chosen neighbor site of the last placed $(n-1)$th  monomer ($n\leq N$). The growth is stopped, 
if the total length of the chain, $N$, is reached (we consider chains of length up to $N=100$).
The total energy of a chain $E(N)$ is given by
\begin{equation}
E(N)=\sum_{i=1}^N\sum_{j=i+2}^{N}\varepsilon_{ij}\sigma_{ij},
\label{senergy}
\end{equation}
where each $i$ and $j$ is either of the type $A$ or $B$, correspondingly $\varepsilon_{ij}$ is $1$, $-1$ or $0$ depending on the model,  and   $\sigma_{ij}=1$ if monomers $i$ and $j$ are nearest neighbors  and zero otherwise.
The Rosenbluth weight factor $W_n$ is thus taken to be
\begin{equation}
W_n=\prod_{l=2}^n m_l {\rm e}^{-\frac{(E(l)-E(l-1) )}{k_B T}},
\label{weight}
\end{equation}
where  $m_l$ is the number of free lattice sites to place the $l$th monomer,   $E(l)$ denotes the energy of the $l$-step chain, and $k_B$ is
the Boltzmann constant.
In what follows, we will assume units in which $k_B=1$.

Pruning and enrichment are 
performed by choosing thresholds $W_n^{<}$
and $W_n^{>}$ depending on the current estimate of the sum of weights  $Z_n=\sum_{{\rm conf}} W_n^{{\rm conf}}$ of the $n$-monomer chain \cite{Grassberger97,Hsu03a,Bachmann03}. 
If the current weight $W_n$ of an $n$-monomer chain is less than $W_n^{<}$,  the chain is discarded with probability 
$1/2$, whereas if $W_n$ exceeds  $W_n^{>}$, the configuration is doubled (enrichment of the sample with high-weight configurations). We 
adjust the pruning-enrichment control such that at least 10 chains are generated per each iteration tour and perform $10^6$ tours.

When studying random hereropolymers, a double averaging has to be performed
to receive the quantitative value for any observable.
At first,  the conformational averaging is performed over an ensemble of possible conformations of a macromolecule with a
fixed sequence of $A$ and $B$ monomers for a given fraction $c$ value along the chain:
\begin{eqnarray}
&&\langle O \rangle  
=\frac{\sum_{{\rm conf}} W_N^{{\rm conf}}O } {\sum_{{\rm conf}} W_N^{{\rm conf}}}, \label{R}
\end{eqnarray}
where  $W_N^{{\rm conf}}$ is the weight of an $N$-monomer chain in a given conformation.
The second is the ``sequence average" $\overline{ \langle ... \rangle}$, which is carried out over 
different random sequences of $A$ and $B$ monomers for a given $c$ value:    
\begin{eqnarray}
&&\overline{\langle O \rangle}=\frac{1}{M}\sum_{i{=}1}^M \langle O\rangle_i. \label{avprob} 
\end{eqnarray}
Here, $M$ is the number of different sequences and the index $i$ means that a given quantity is calculated for sequence $i$. 
Note, that the number of all possible sequences of an $N$-monomer chain at fixed fraction $c$ of $N_A$ monomers, given by
$$M=\frac{N!}{(cN)!(N-(cN)!)},$$ becomes incredibly large with growing of $N$, especially with $c$ close to $0.5$.  
In our calculations, we consider up to $M=2500$ random sequences at given fraction $c$;  
this, however, leads to reasonable results as is shown by comparison with some phenomenological predictions (see below). 
Note, that the case of so-called ``quenched disorder'' is considered, 
where the average over different sequences is taken after the configurational average has been performed.

\begin{figure}[t!]
\includegraphics[width=5.4cm]{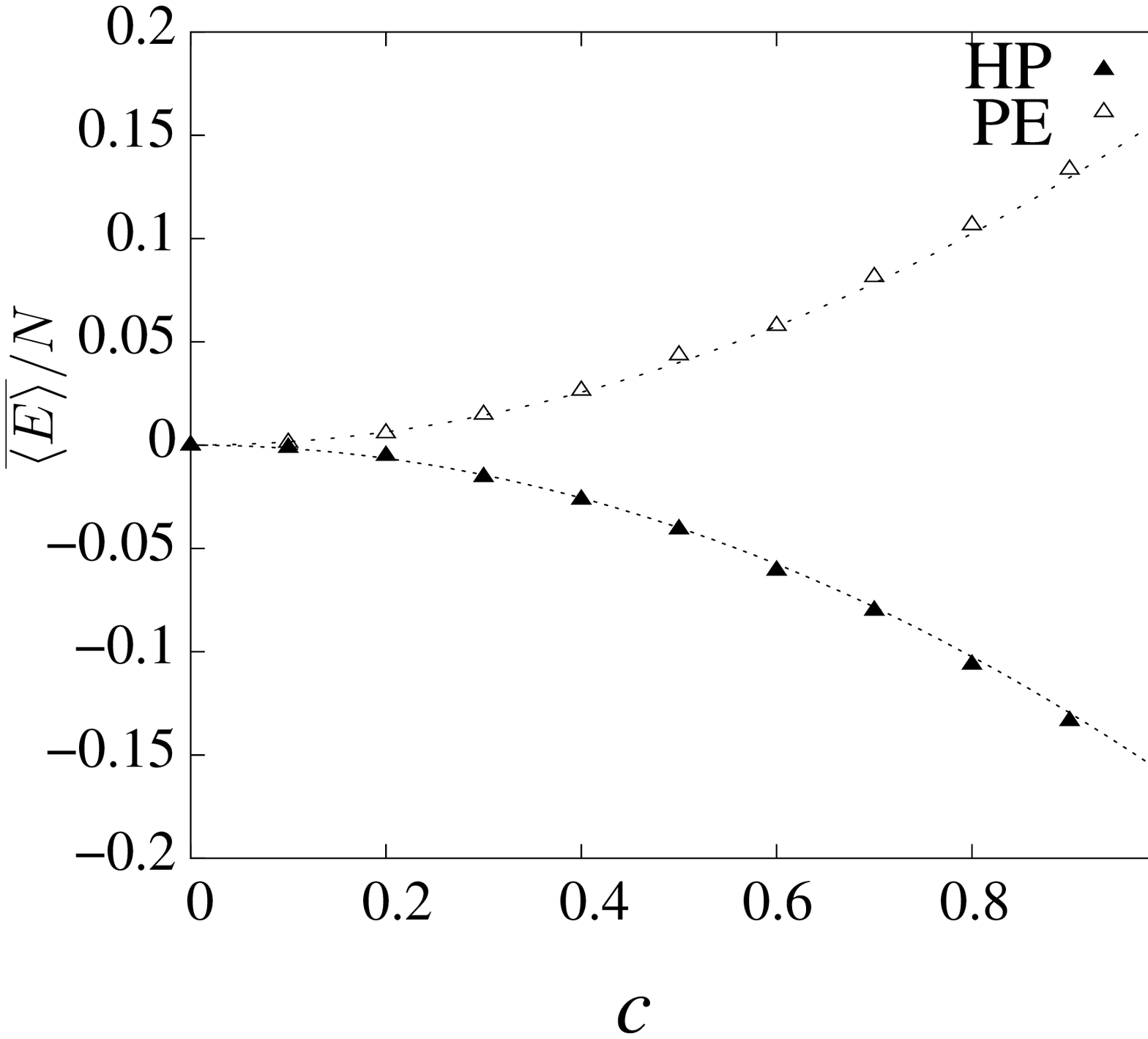}
\hspace*{0.2cm}
\includegraphics[width=5.4cm]{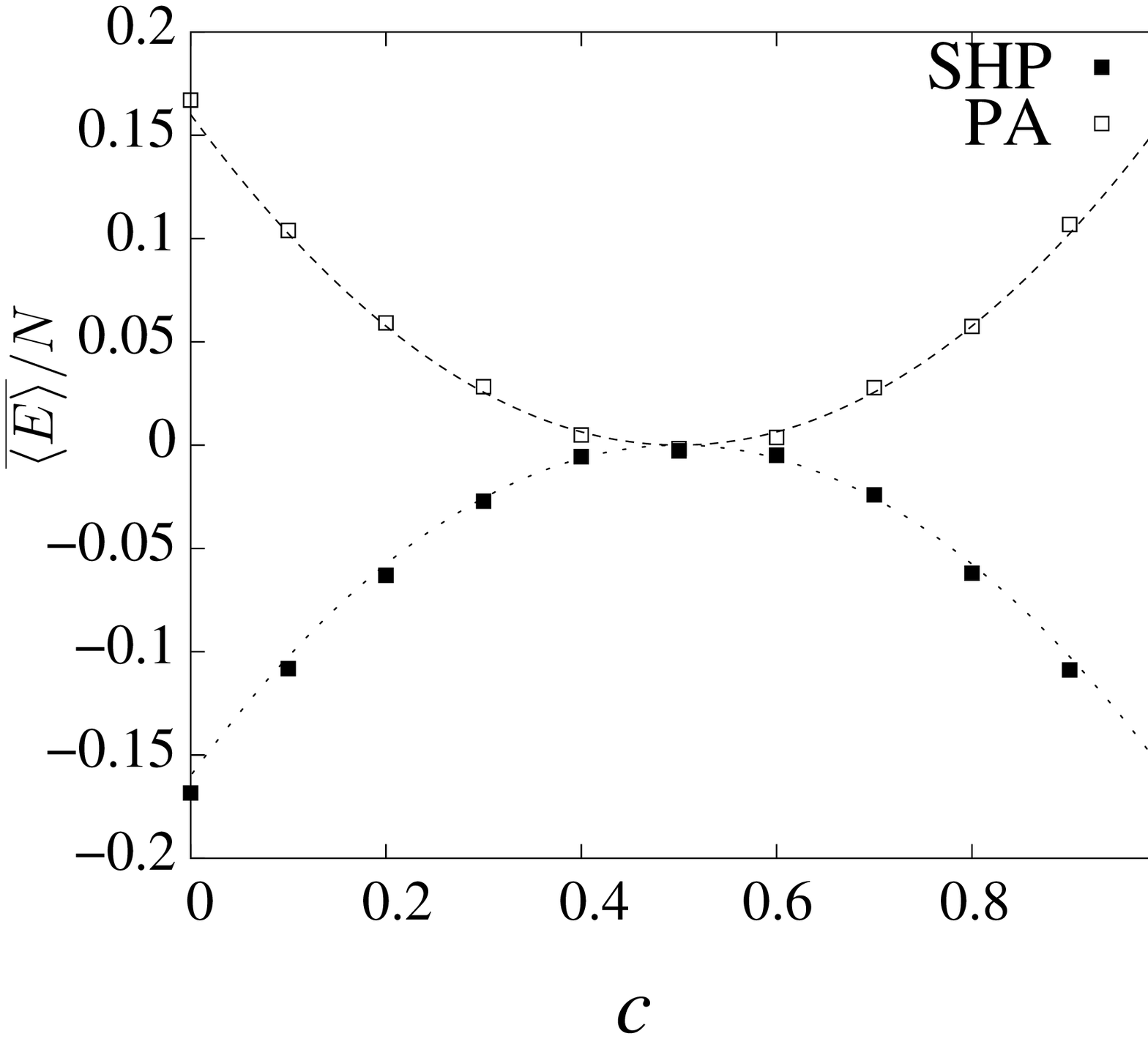}
\hspace*{0.2cm}
\includegraphics[width=5.4cm]{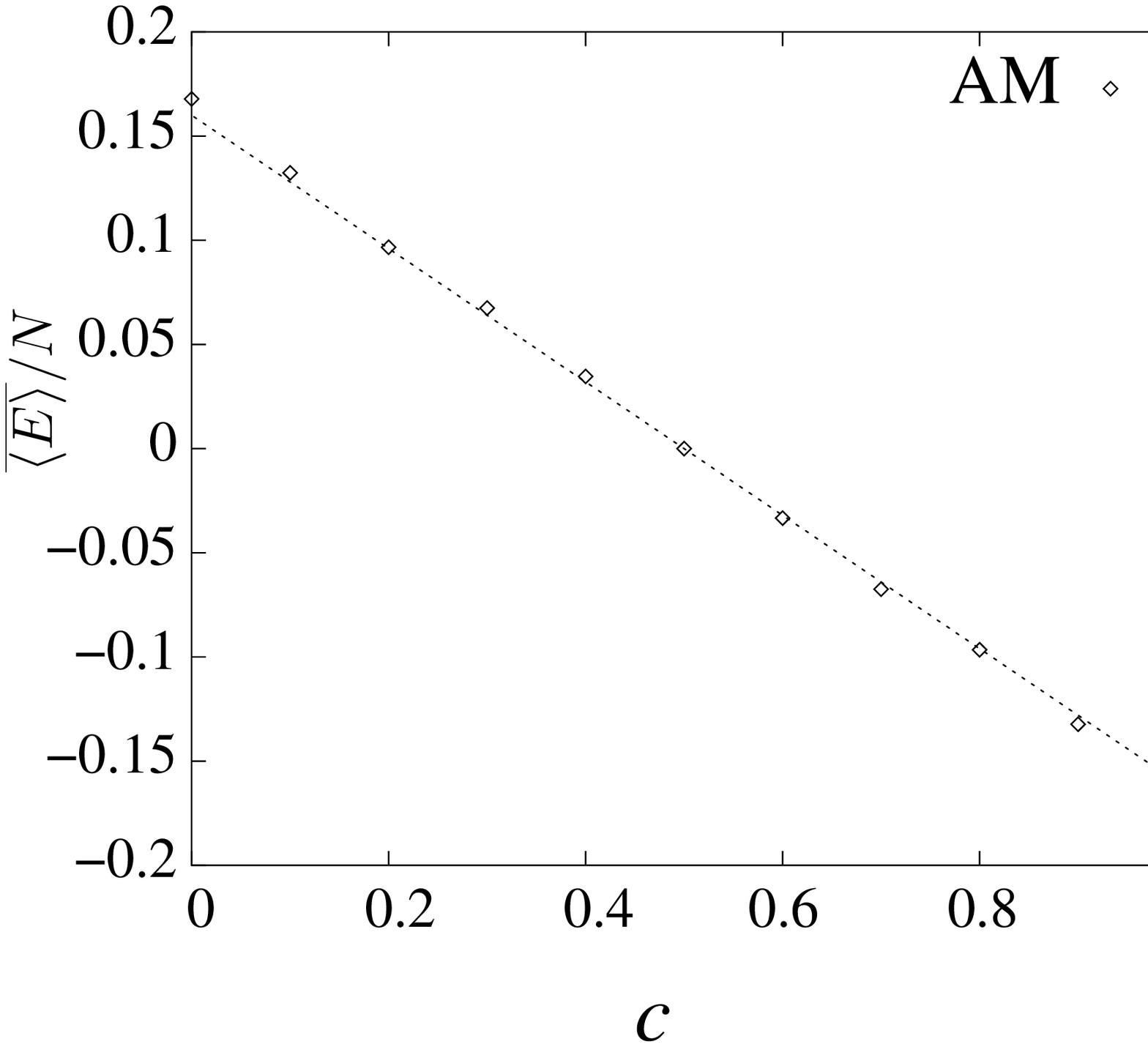}
\caption{\label{encoil} The averaged energy per monomer of an $N=100$-monomer heterogeneous chain in the high-temperature
 regime with various types of intramonomer interaction as function of $c$. 
Lines are the corresponding results of empirical estimations (\ref{en1})-(\ref{en5}).}
\end{figure}

\section{Results}

Estimates for the averaged energy of heteropolymer models (\ref{mod1})-(\ref{mod5}) can be obtained based on the following
considerations. 
Let us introduce the averaged number of nearest-neighbor contacts $p(d)$ for any monomer of a chain, which in $d$ dimensions can be estimated from the empirical relation: 
\begin{equation}
z(d) = 2d - 1 - p(d),\label{nn}
\end{equation}
where $z(d)$ is the connectivity constant (averaged number of possibilities to make the next step in a growing SAW trajectory).
For an idealized Gaussian chain, when the trajectory is allowed to cross itself,
 for the case of a regular $d$-dimensional lattice one has $z(d)=2d$. Taking into account self-avoidance effect, one first of all forbids the ``turning back'' of the 
 trajectory at each step, which reduces the connectivity constant to the value $2d-1$. Finally, another factor which reduces the connectivity constant at each step due to the self-avoidance effect, 
 is the number of nearest sites, already visited by the trajectory (number of contacts with nearest neighbors $p(d)$). 
Having a heterogeneous polymer chain in $d=3$ with fixed fraction $c$ (so that $N_A=N\cdot c$, $N_B=N\cdot(1-c)$), we can estimate
the number of  $AA$ nearest-neighbor contacts:
\begin{equation}
n_{AA}=\frac{1}{2}N_A p(d=3)  c=\frac{1}{2}N p(d=3)  c^2,
\end{equation}   
the number of  $BB$ nearest-neighbor contacts:
\begin{equation}
n_{BB}=\frac{1}{2}N_B p(d=3)  (1-c)=\frac{1}{2}N p(d=3) (1-c)^2,
\end{equation}  
and the number of  $AB$ nearest-neighbor contacts:
\begin{equation}
n_{AB}=N_Ap(d=3)(1-c)=Np(d=3)c(1-c).
\end{equation} 
Thus, for the total energies of models (\ref{mod1})-(\ref{mod5}) we have correspondingly:
\begin{eqnarray}
&&E_{{\rm  HP}}=- n_{AA}=-\frac{1}{2}N p(d=3)  c^2,\label{en1}\\
&&E_{{\rm SHP}}=-n_{AA}-n_{BB}+n_{AB}=-2N p(d=3) \left(c^2-c+\frac{1}{4}\right),\label{en2}\\
&&E_{{\rm PA}}=n_{AA}+n_{BB}-n_{AB}=2N p(d=3) \left(c^2-c+\frac{1}{4}\right), \label{en3}\\
&&E_{{\rm  PE}}= n_{AA}=\frac{1}{2}N p(d=3)  c^2,\label{en4}\\
&&E_{{\rm  AS}}=-n_{AA}+n_{BB}= Np(d=3)\left(\frac{1}{2}-c\right)\label{en5}.
\end{eqnarray}

In the high-temperature regime,  the steric self-avoidance effect between monomers plays the only role in determining
the conformational properties of polymers. Thus  taken the known value $z(d=3){=}4.68404(9)$ \cite{Clisby} (which gives
$p(d=3){=}0.31596(9)$ according to (\ref{nn})), we can obtain numerical  estimates for (\ref{en1})-(\ref{en5}). 
These expressions perfectly agree with results of our  numerical simulations, see Fig.~\ref{encoil} where the corresponding energies per 
 monomer are presented.

\begin{figure}[t!]
\includegraphics[width=5.4cm]{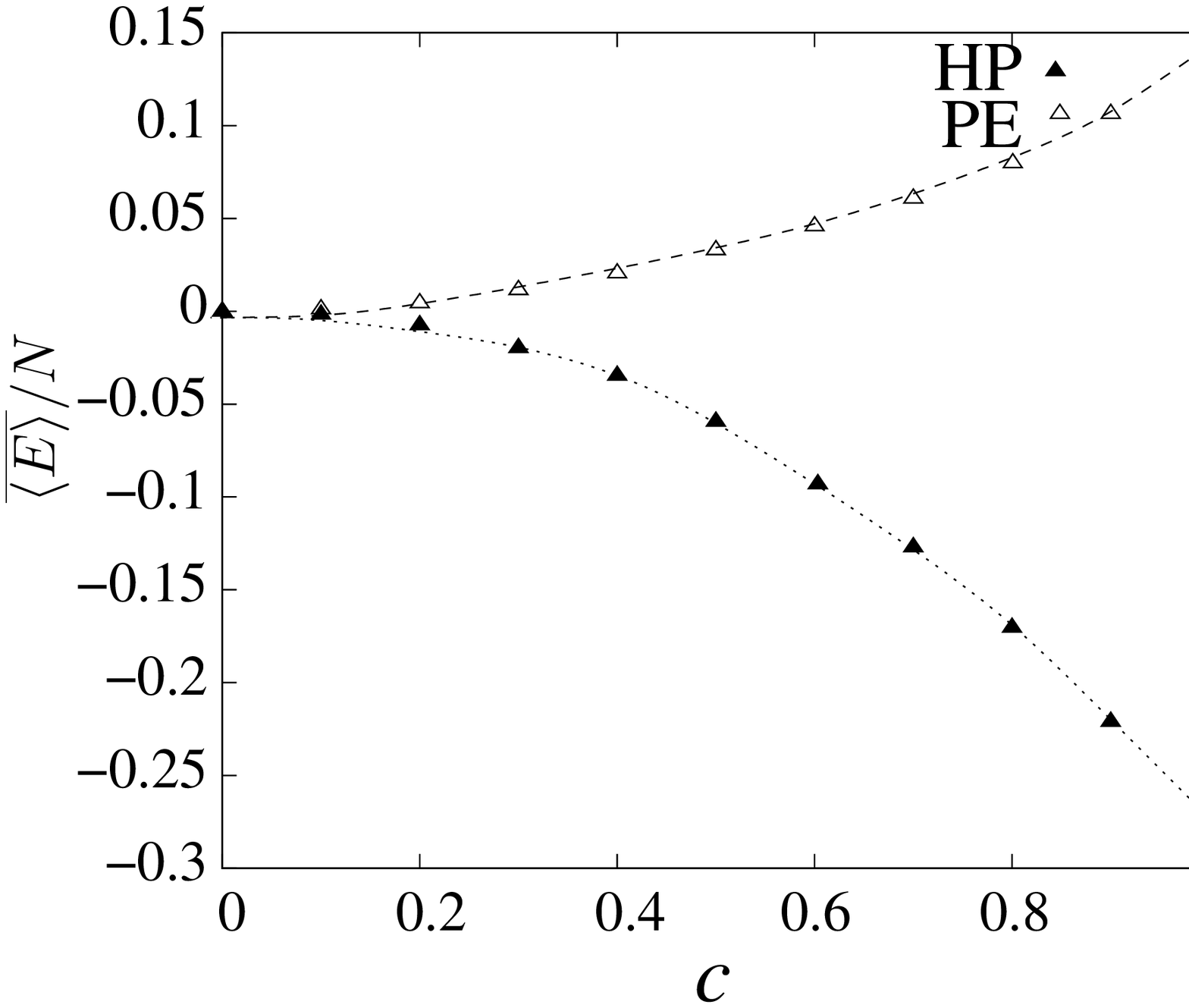}
\hspace*{0.2cm}
\includegraphics[width=5.4cm]{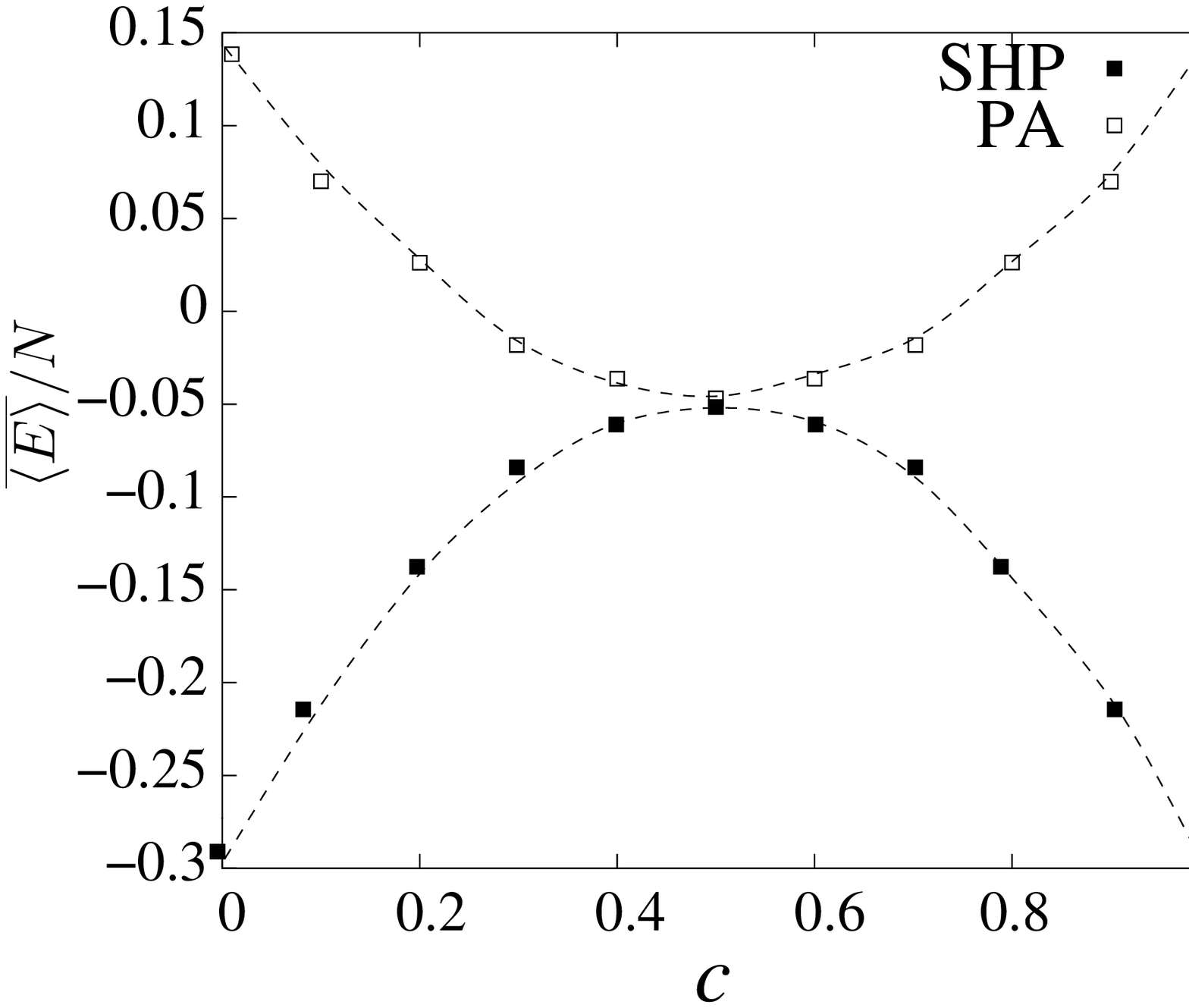}
\hspace*{0.2cm}
\includegraphics[width=5.8cm]{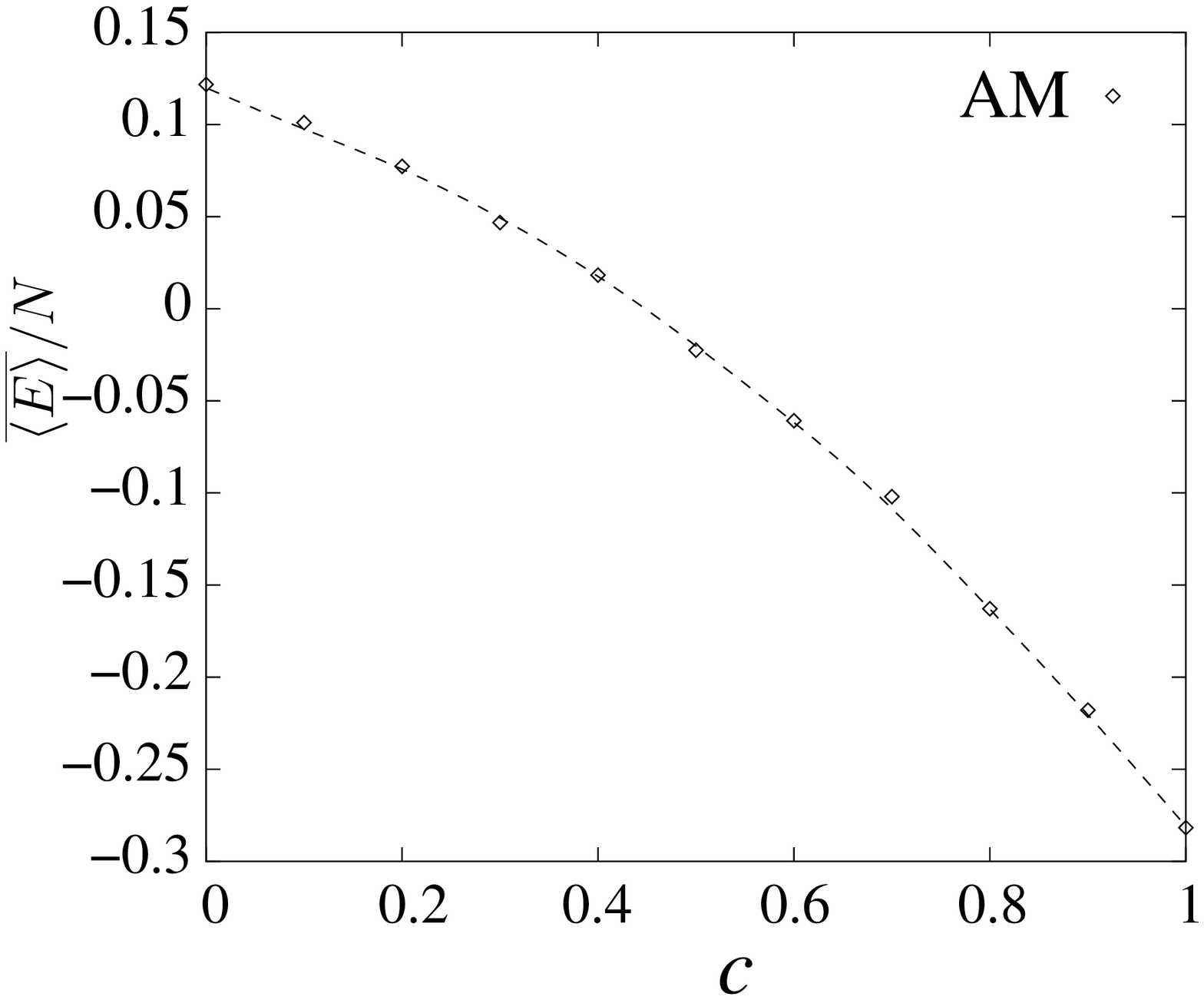}
\caption{\label{encoil_T} The averaged energy per monomer of an $N=100$-monomer heterogeneous chain  with 
various types of intramonomer interaction as function of $c$ at temperature $T=4.0$. Lines are guides to the eye.}
\end{figure}

With lowering the temperature, the interactions between monomers become essential, and the $p$-values for each model will depend on 
the fraction $c$ and temperature $T$. As an example, we present our simulation results for the corresponding energies per 
 monomer at $T=4.0$ in Fig.~\ref{encoil_T}. Though for all models the polymer chains are expected to be still in an extended state 
(coil regime) at this temperature, the typical conformations they 
attain at given $c$ are governed by that nearest-neighbor contact value which minimizes the total 
energy. Indeed, for the polyelectrolye model (\ref{mod4}) the most preferable configurations would be those with a small value of nearest-neighbor contacts,
whereas for the  HP model (\ref{mod1}) the largest possible value of $p$ will minimize the total energy.   
The values of $p$ for all the models at various $c$-values can be estimated on the basis of data presented in Fig. \ref{encoil_T}
applying expressions (\ref{en1})-(\ref{en5}). 

\begin{figure}[b!]
\begin{center}
\includegraphics[width=7.0cm]{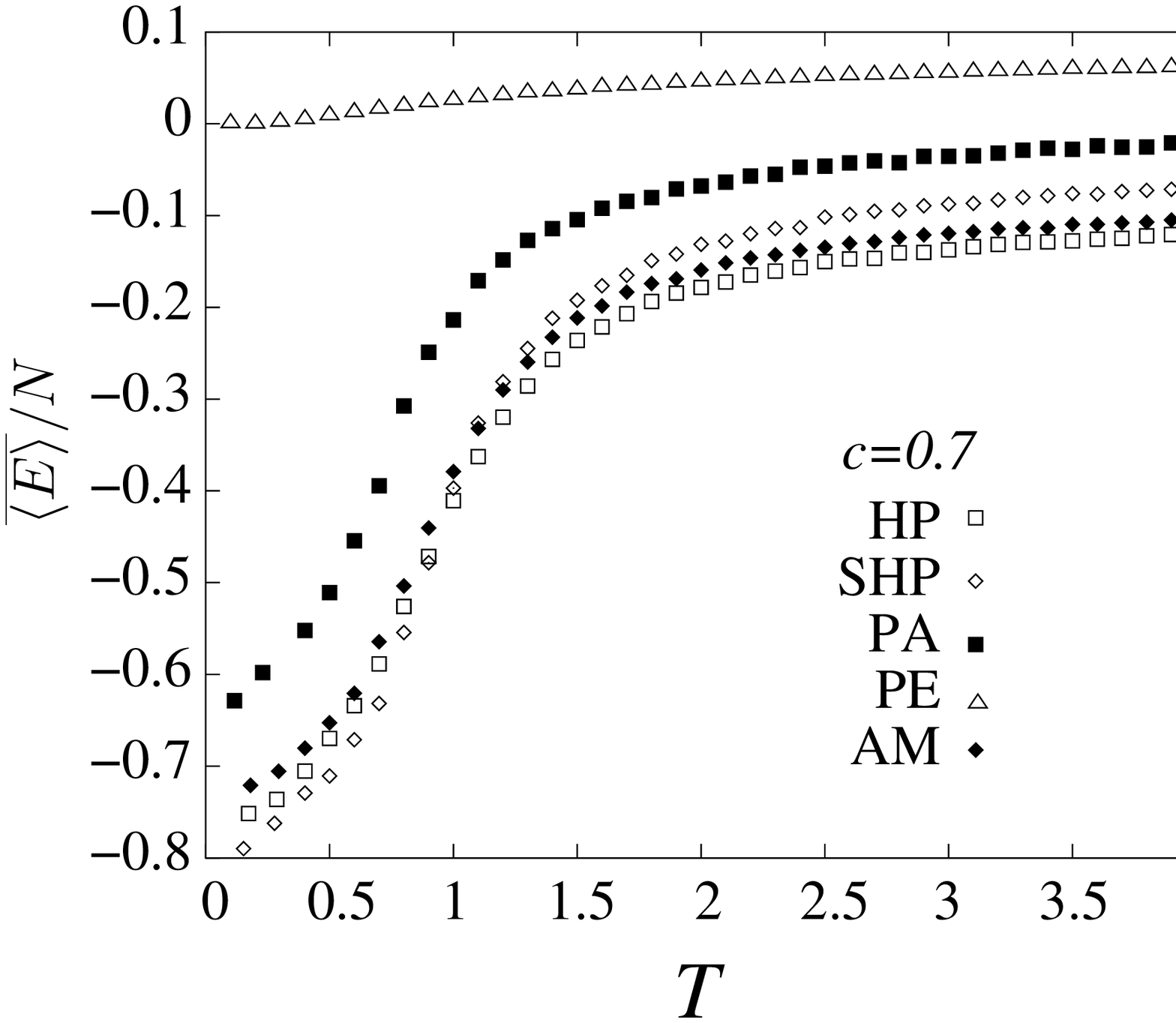}
\hspace*{1cm}
\includegraphics[width=6.9cm]{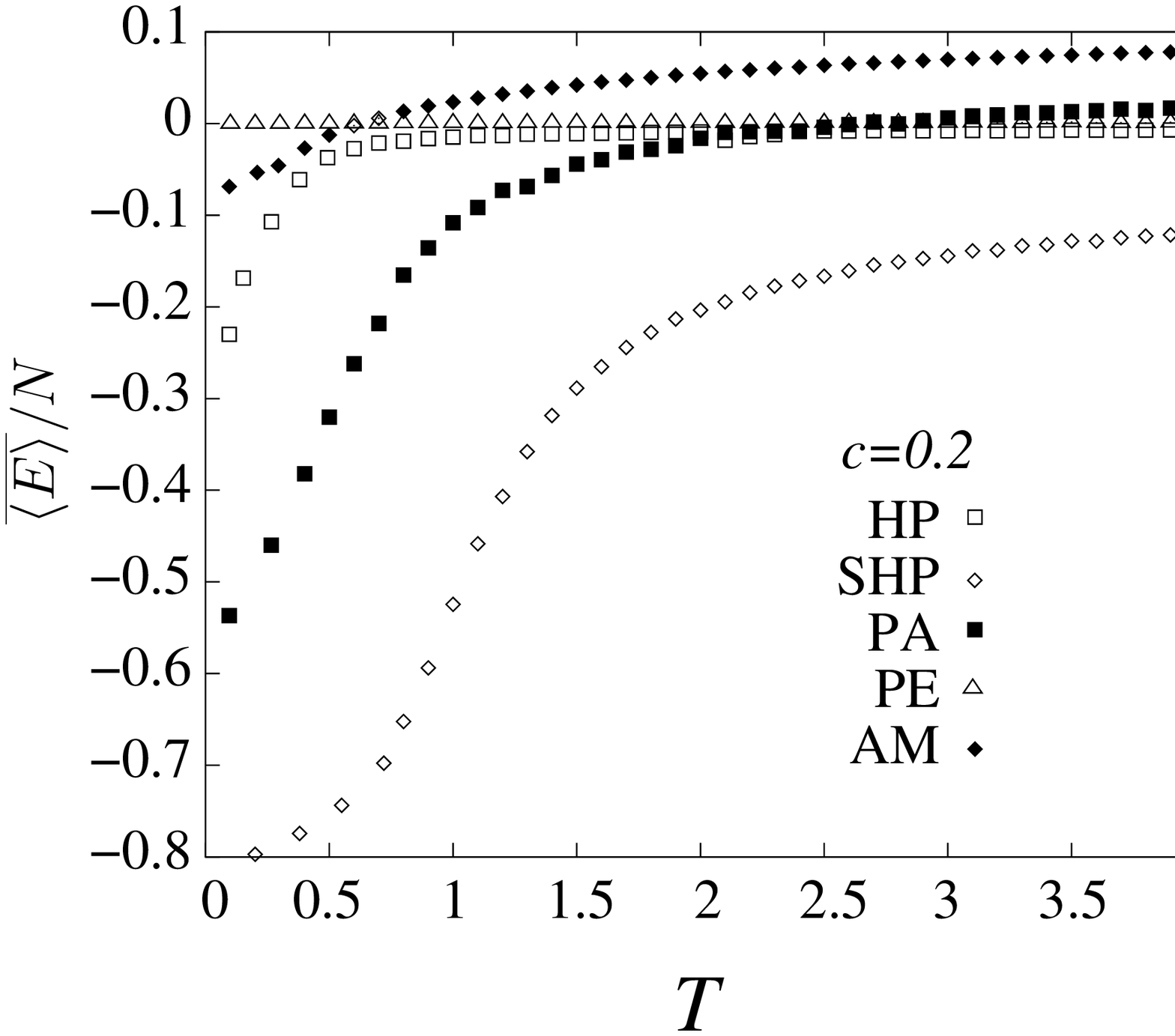}
\end{center}
\caption{\label{ener_total} Averaged energy per one monomer of the heterogeneous polymer chain
 with various types of monomer-monomer interactions as function of temperature at fixed fraction $c=0.7$ and $c=0.2$.}
\end{figure}

In Fig. \ref{ener_total} we present our results for the averaged energies of the $N=100$-monomer heterogeneous chain as function of temperature 
at two representative fixed fractions $c=0.7$ and $c=0.2$. At $c=0.7$, the energy of the models $\ref{mod1}$, $\ref{mod2}$, $\ref{mod3}$ and $\ref{mod5}$  
decreases monotonically as temperature is lowered and
 sharply falls down within a short temperature interval, indicating the transition from an extended state (coil regime) to a compact state. 
 For the polyelectolyte model (\ref{mod4}), however, 
one observes only a monotonic decrease of energy, which gradually tends to zero with lowering the temperature, and thus no transition occurs. Moreover, 
within the frames of this model, decreasing the energy means decreasing  the averaged number of nearest-neighbor  contacts, 
and thus expanding the chain. Thus, unlike the other models, in this case the averaged size of the heteropolymer chain increases with lowering the temperature. 
Similar results are found at fixed fraction $c=0.2$. In this case, model (\ref{mod5})
consists mainly of monomers, which repel each other, and thus behaves like the polyelectrolyte (\ref{mod4}):  conformations with a minimum  number of 
nearest-neighbor  contacts are energetically preferable, which leads to expanding the chain size with lowering the temperature. 
Note also that at small concentration of attractive monomers, the chains can attain the compact state only when they are long enough 
and have enough attractive nearest-neighbor contacts to overcome the conformational entropy.

\begin{figure}[t!]
\includegraphics[width=7.1cm]{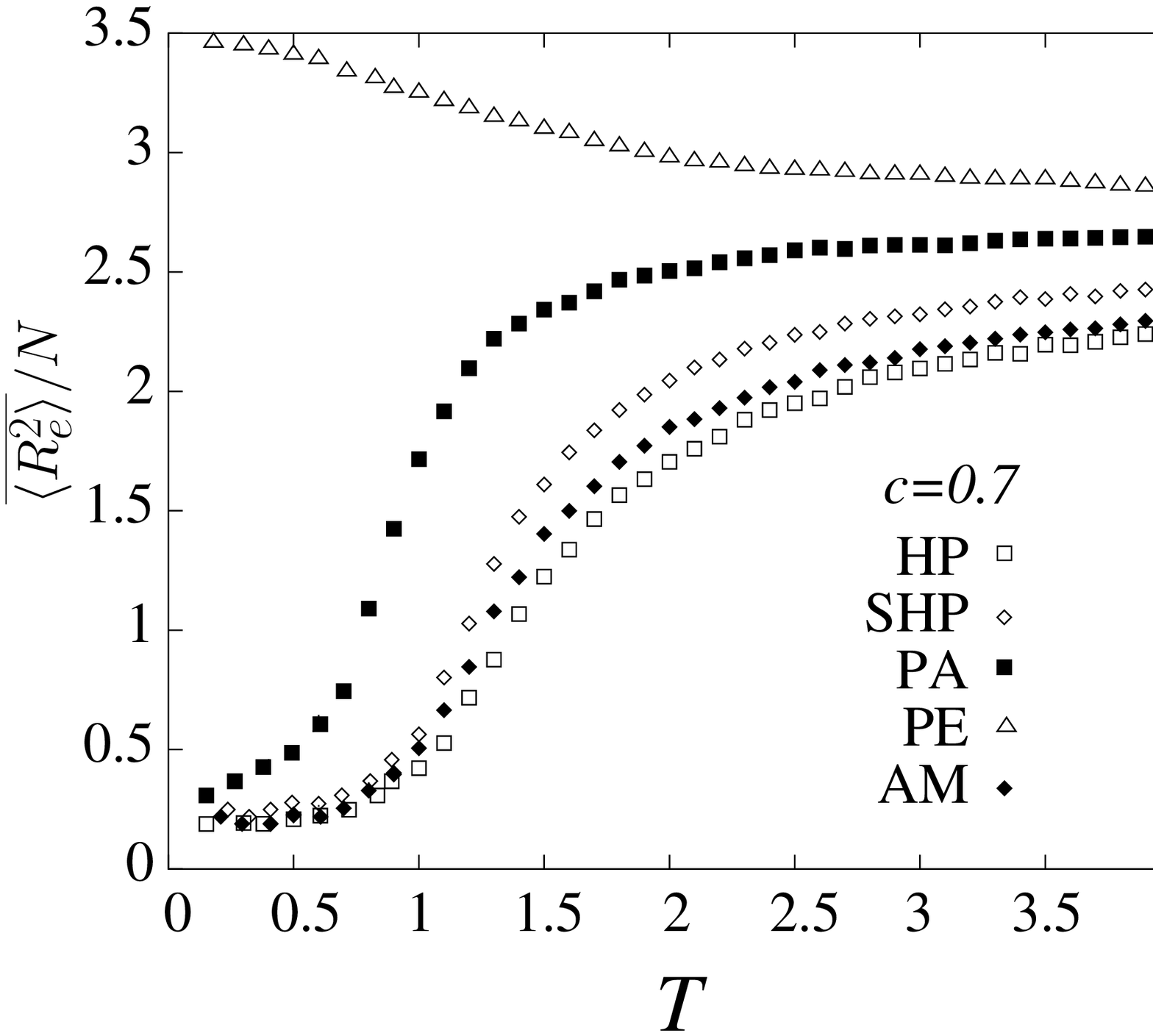}
\hspace*{0.5cm}
\includegraphics[width=7.1cm]{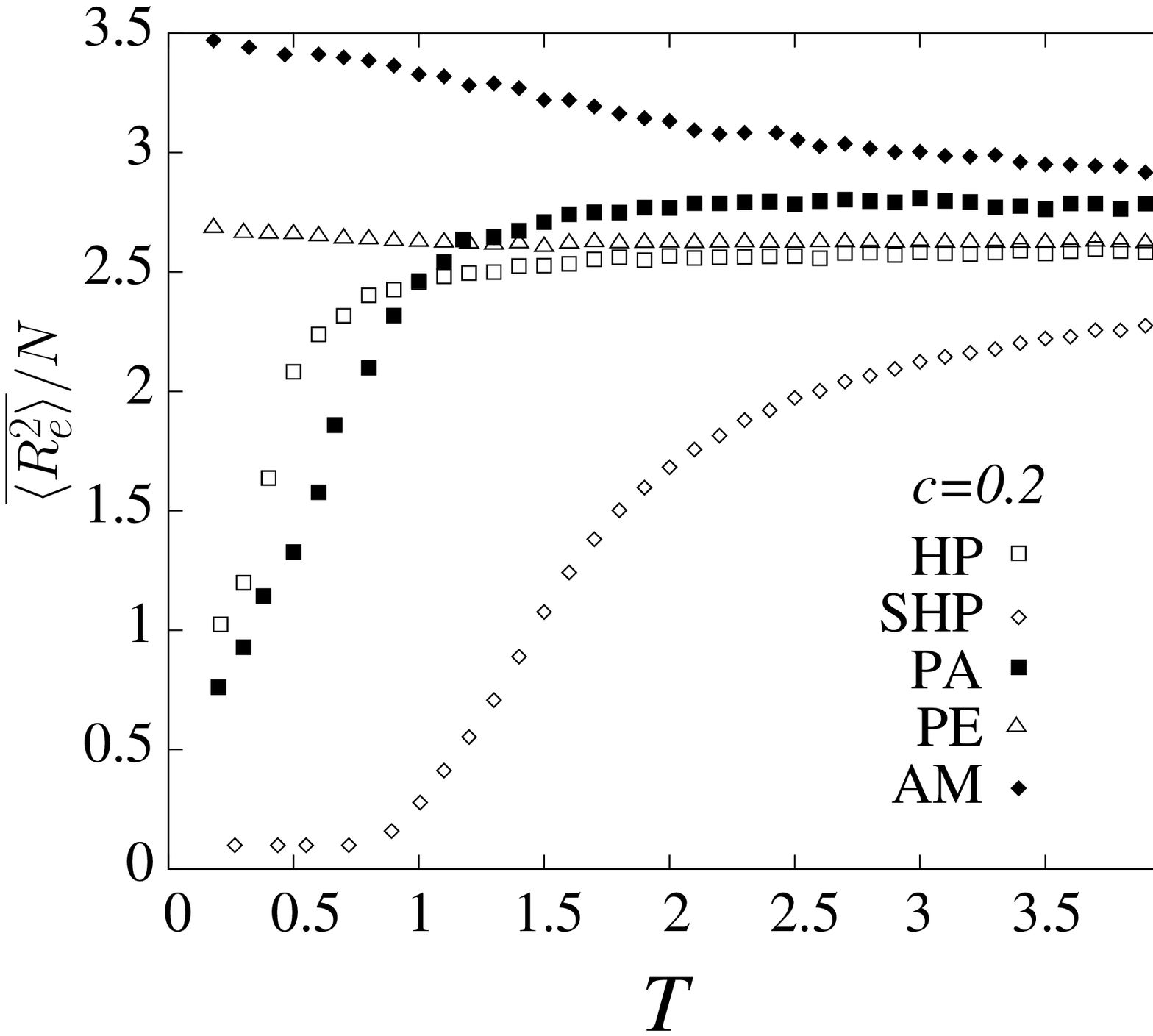}
\caption{\label{R_total} Averaged end-to-end distance of the heterogeneous polymer chain with various types of monomer-monomer interactions as function of temperature at fixed fraction $c=0.7$ and $c=0.2$. }
\end{figure}

In Fig.~\ref{R_total} our results for the averaged end-to-end distance of the $N=100$-monomer heterogeneous chain as function of temperature are given at the
same two representative fractions $c=0.7$ and $c=0.2$. These results confirm our discussions above. For the case $c=0.7$, one notices
a decrease of the polymer chain size with lowering the temperature (models $\ref{mod1}$, $\ref{mod2}$, $\ref{mod3}$ and $\ref{mod5}$),
which becomes crucial close to the
 transition temperature into the compact state. However, 
for the polyelectrolyte (\ref{mod4}) the polymer chain is expanding its size  with lowering the temperature due to repulsion between monomers, and the 
polymer chain remains
in an extended state at any temperature. At  $c=0.2$, the size of the polymer chain of model (\ref{mod5}) increases even stronger than that of a polyelectrolyte due to the considerably larger amount of mutually repelling monomers; in both cases the polymer 
chain remains in an extended state at any temperature. 

\begin{figure}[b!]
\begin{center}
\includegraphics[width=7.3cm]{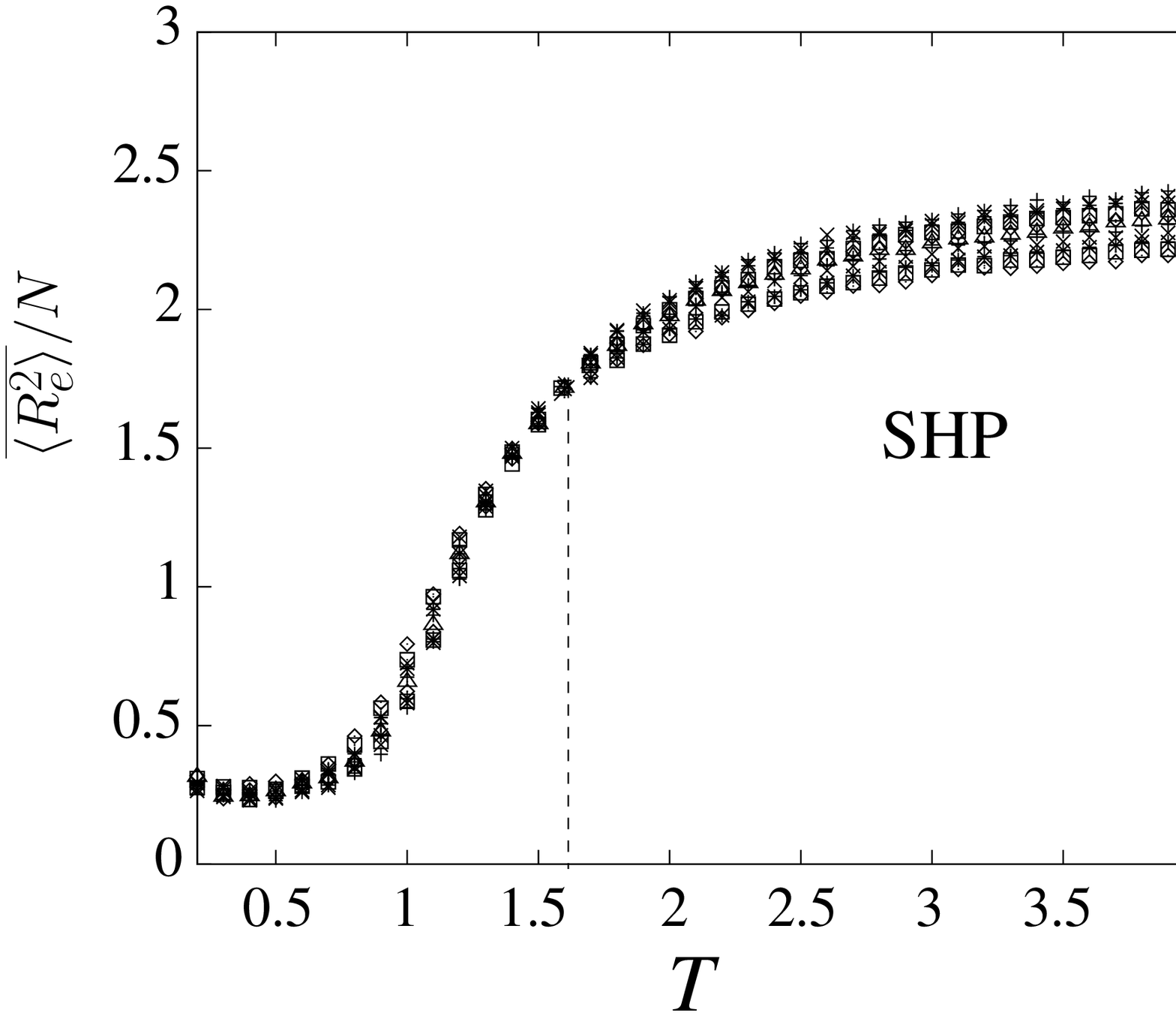}
\hspace*{0.5cm}
\includegraphics[width=7.3cm]{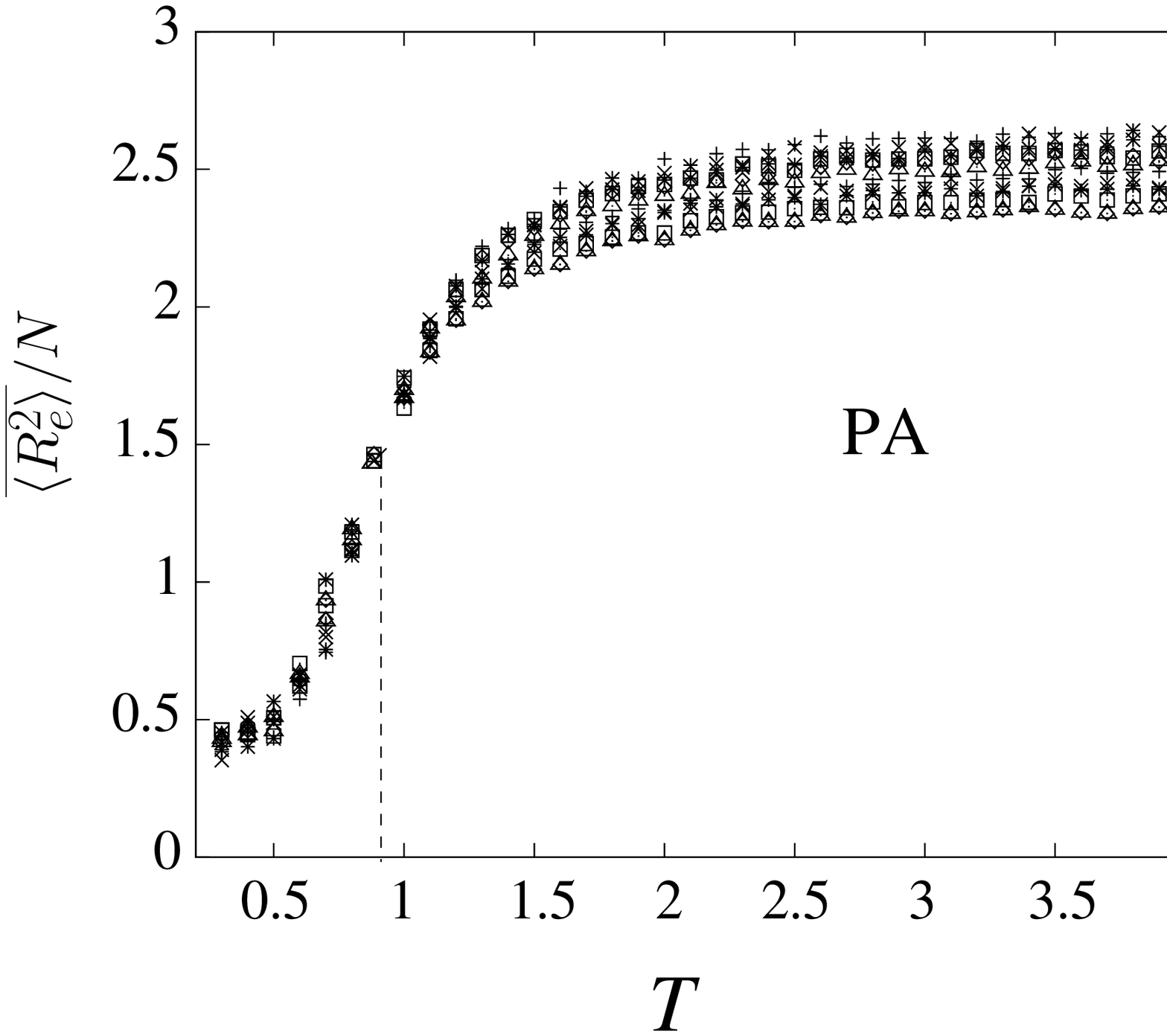}
\end{center}
\caption{\label{cros} Averaged end-to end distance of a heterogeneous polymer chain  divided by number of monomers as function of temperature 
at fraction $c=0.7$ at various values of $N$ from 68 up to 100.}
\end{figure}

To obtain a  quantitative description of conformational transitions from extended to compact states in models (\ref{mod1})-(\ref{mod5}), 
we study the ratio  $\overline{\langle R_e^2 \rangle}/N$ for various $N$ and fixed fraction $c$ 
as function of temperature (Fig. \ref{cros}).
Assuming that the collapse of heteropolymers is through the $\Theta$-transition similar to 
homogeneous polymers, we expect $\overline{\langle R_e^2 \rangle}\sim N$ at the $\Theta$-point,
and thus the curves intersect in the vicinity of the transition temperature (the larger $N$, the steeper the curves). Despite the slight dependence of 
the intersection points on $N$, we can locate the crossover temperature. 
If at some $T$ and $c$ values the curves do not intersect, there is no phase transition and the polymer is in an extended state
at any temperature.

\begin{figure}[t!]
\includegraphics[width=6cm]{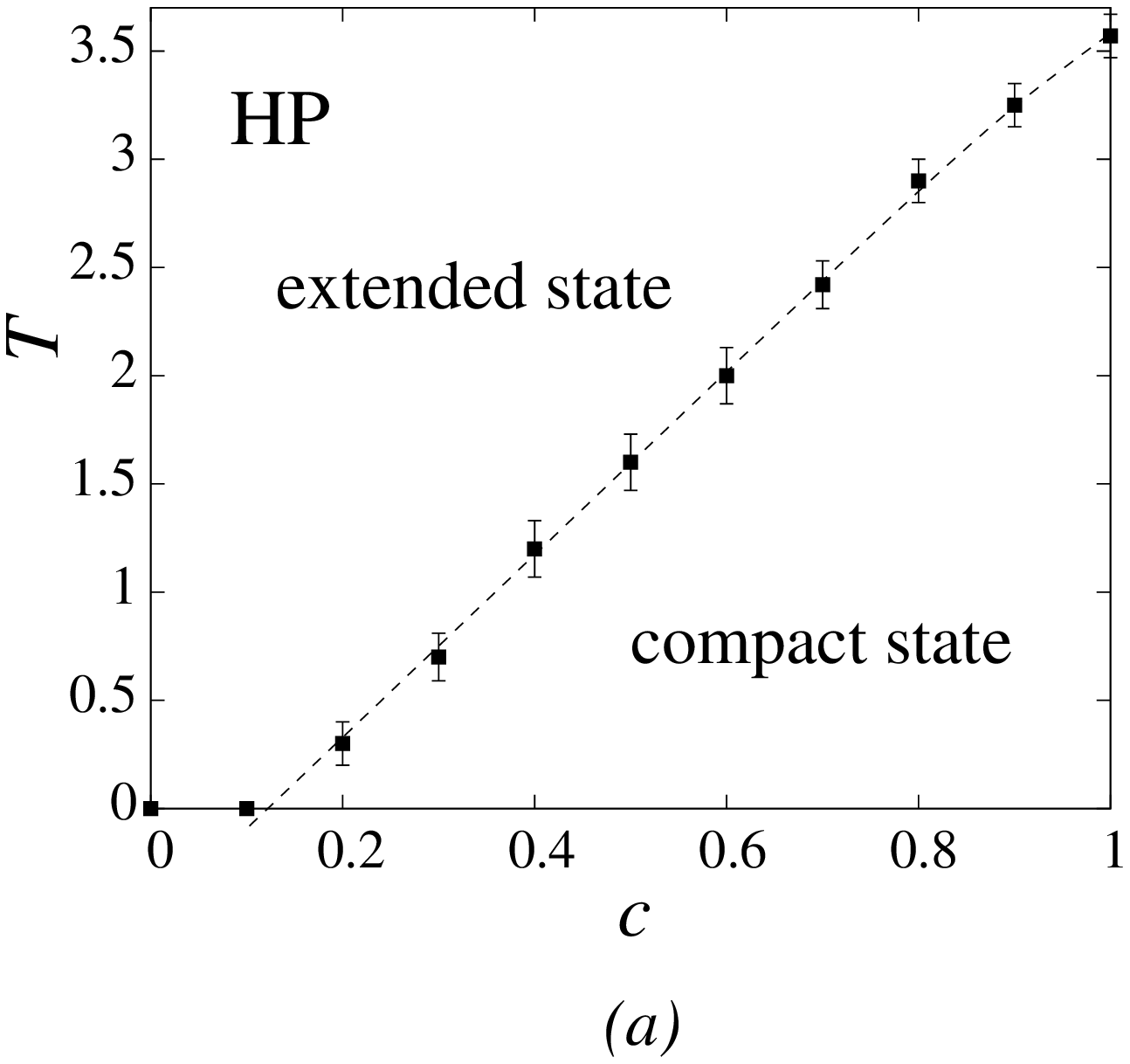}
\includegraphics[width=6cm]{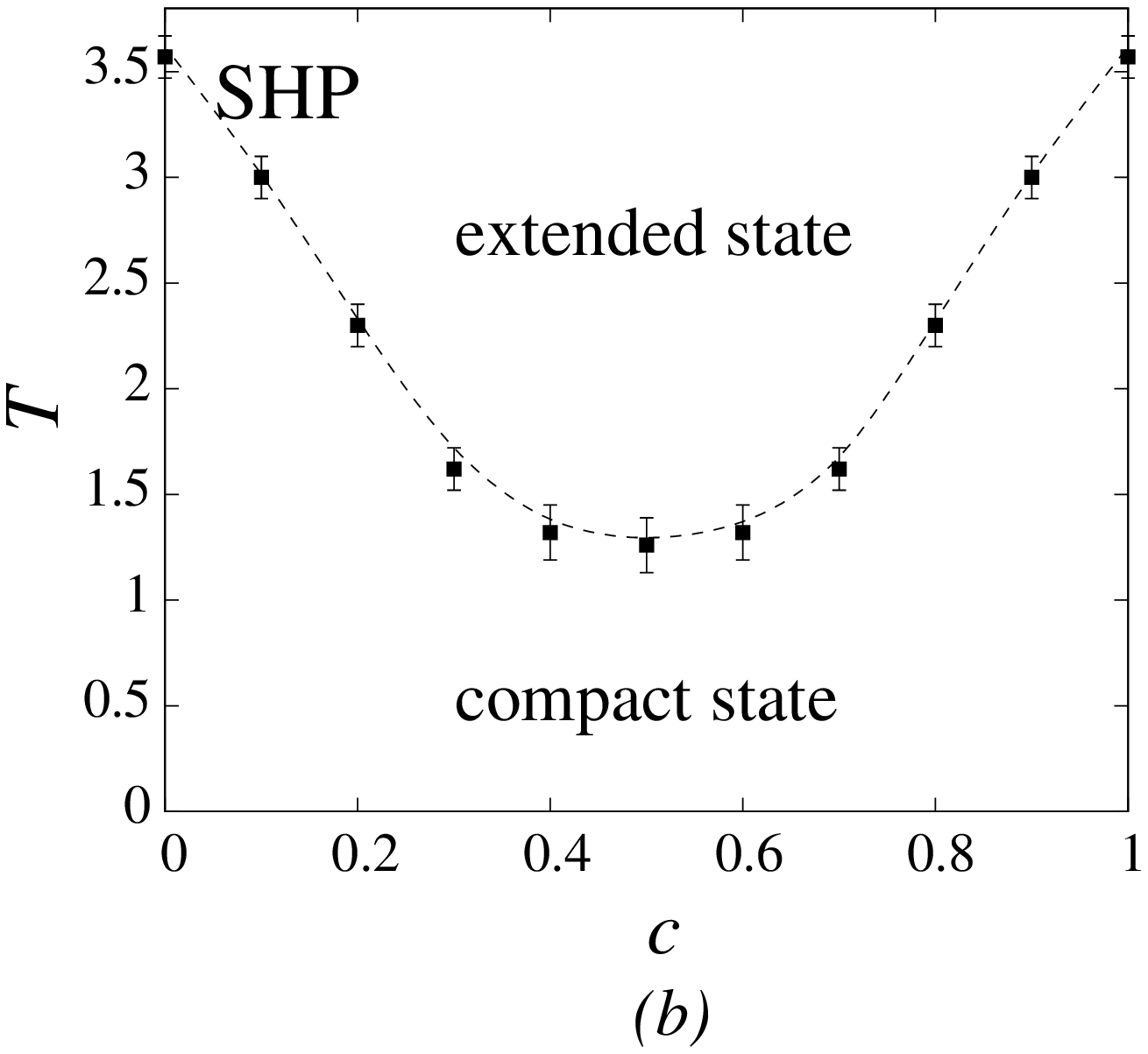}
\includegraphics[width=6cm]{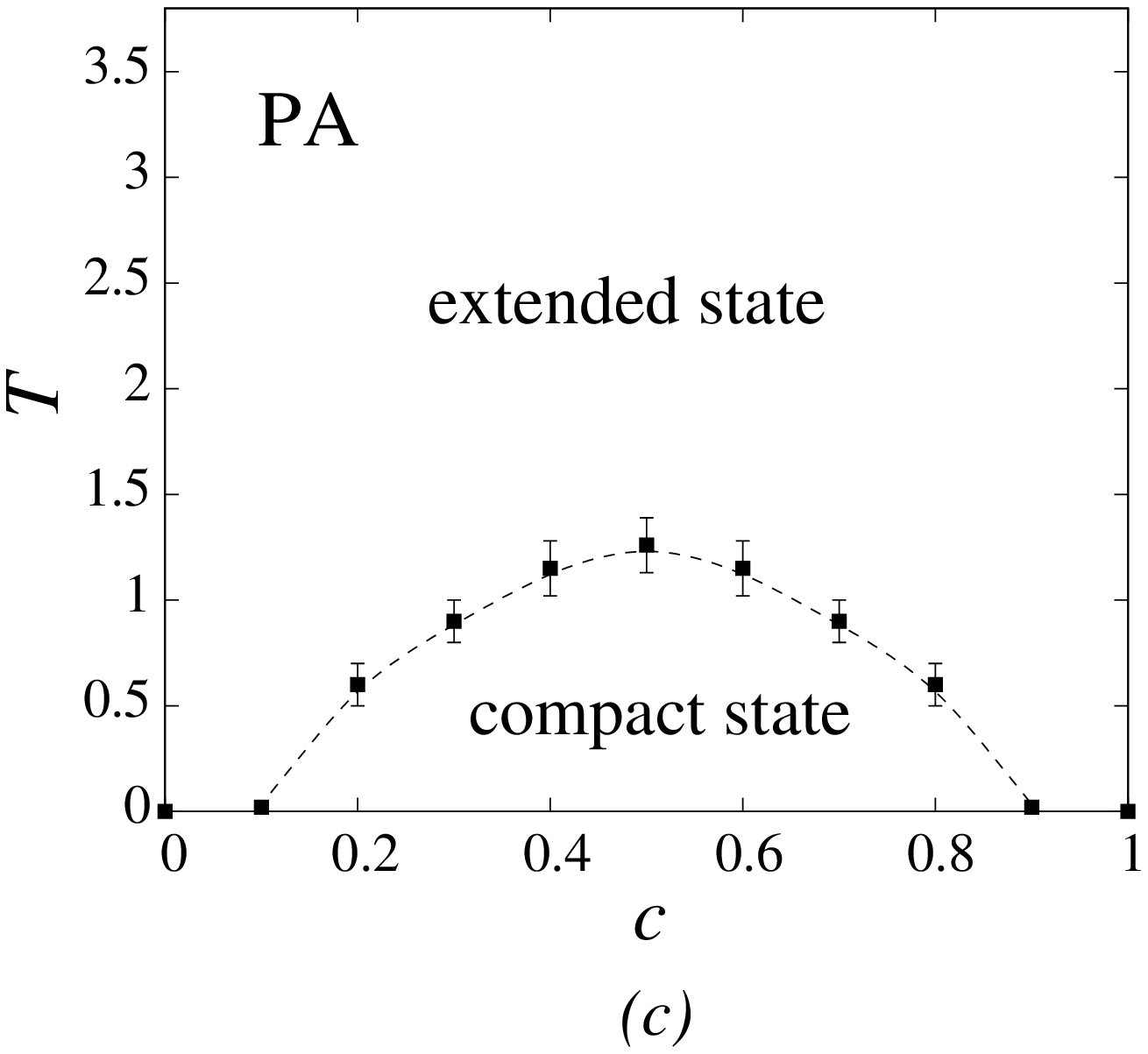}
\includegraphics[width=6cm]{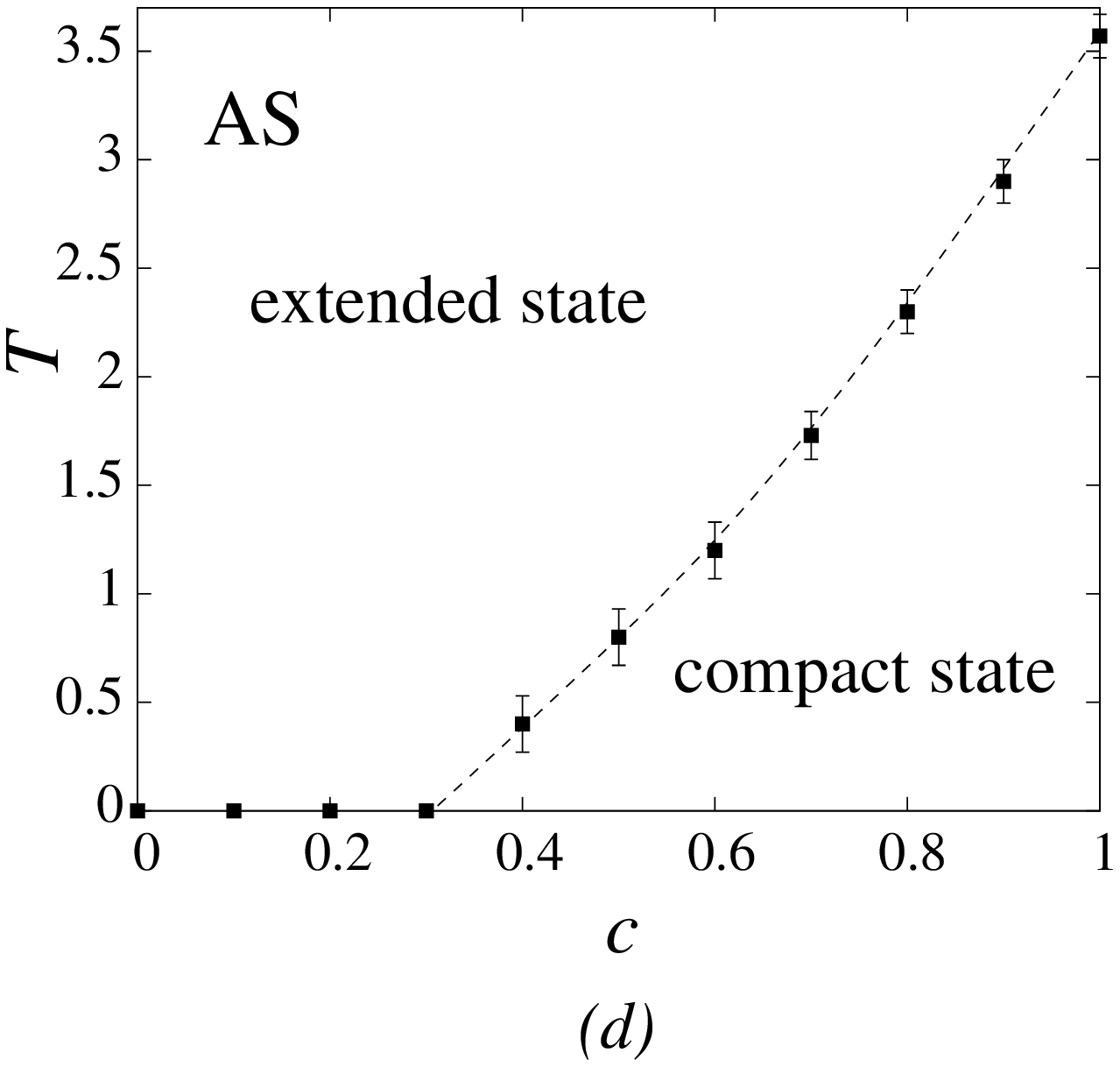}
\caption{\label{diagram}Phase diagrams of heterogeneous polymer chain in $T$-$c$ space. (a): model (1), (b): model (2), (c): model (3), (d): model (5). The polyelectrolyte model (\ref{mod4})
always stays in the extended state and does not exhibit any phase transition.}
\end{figure}

As a result, we obtain the phase diagrams, presented in Fig. \ref{diagram}. 
Note, that within the frames of polyelectolyte model (\ref{mod3}) the polymer chain is always in an extended state 
at any temperature and fraction $c$ value,
thus we do not construct any phase diagram for this model. In the 
limiting case $c=1$, the  models (\ref{mod1}), (\ref{mod2}) and (\ref{mod5}) 
describe homogeneous  polymer chains with nearest-neighbor attractions (for model 2 also $c=0$) with known value of transition temperature $T_{\Theta}=3.717(3)$ \cite{Grassberger97}.
For the symmetrized HP model (\ref{mod2}), the $\Theta$-transition is present always at any value of $c$, 
whereas the other models remain in the extended state when 
the concentration of attracting monomers is too small to cause a transition into the compact state. The results for the polyampholyte model (\ref{mod3}) are in a good agreement with those obtained previously in Ref. \cite{Kantor94}.


\section{Conclusions}

We studied the conformational properties of heteropolymers in $d=3$ dimensions within the frames of 
a lattice model containing  $N_A$ monomers of type $A$ and $N_B=N-N_A$ monomers of type $B$. 
Restricting ourselves  to only short-range interactions between any pair of monomers residing on neighboring lattice sites
that are not connected by a covalent bond, we considered several  generalizations (\ref{mod1})-(\ref{mod5}) of this model. 
In particular,  model (\ref{mod1}) refers to the (minimal) HP model \cite{Lau89} with hydrophobic ($A$) and hydrophilic ($B$) monomers, 
(\ref{mod2}) is the standard model for copolymers with monomers that have a tendency to segregate \cite{Sfatos97}, 
cases (\ref{mod3}) and (\ref{mod4}) stand correspondingly for polyampholytes  \cite{Kantor94} and polyelectrolytes \cite{Tanford61,Dobrynin04} with 
strongly screened Coulomb interactions. Case (\ref{mod5}) 
can be considered as additional generalization of above models, comprising an interplay between different kinds of short-range interactions.

In the present work, we extended the picture, developed in Ref. \cite{Kantor94} for the polyampholyte model (\ref{mod3})  to the whole 
range of cases (\ref{mod1})-(\ref{mod5}); namely, we treat different sequences of $A$ and $B$ monomers at given fixed fraction $c$ (ranging from 0 to 1)
as different realizations of quenched disorder, and obtained observables of interest after performing double conformation and sequence averaging. 
Applying the pruned-enriched Rosenbluth chain-growth algorithm (PERM) we analyzed numerically the peculiarities of transitions 
from the extended into the compact state as a function of the fraction $c\equiv N_A/N$ for all the heteropolymer chain models, 
and obtained the diagrams of phase coexistence in $c$-$T$ space. 

For the polyelectrolyte model (\ref{mod4}),  unlike the other cases, the polymer chain is expanding its size 
with lowering the temperature due to repulsion between monomers, and the 
polymer chain remains in an extended state at any temperature. 
In the symmetrized HP model (\ref{mod2}), the $\Theta$-transition is present always at any value of the fraction $c$, whereas the
other models remain in an extended state when the concentration of attracting monomers is too small to cause a transition into the compact state. 
Note also that at small concentration of attractive monomers, the chains can attain the compact state only when they are long enough 
and have enough attractive nearest-neighbor contacts to overcome the conformational entropy.  
In the limiting case $c=1$, the HP model (\ref{mod1}), the symmetrized HP model (\ref{mod2}) and  model (\ref{mod5}) describe homogeneous  polymer 
chains with nearest-neighbor attractions  (for model 2 also $c=0$) 
with known value of the transition temperature $T_{\Theta}=3.717(3)$ \cite{Grassberger97}.

\section*{Acknowledgements}
This work was supported by an Institute Partnership grant of the 
 Alexander von Humboldt Foundation. W.J. acknowledges support by 
 DFG Sonderforschungs\-bereich SFB/TRR 102 (Project B04) and 
 S\"achsische DFG Forschergruppe FOR877 under Grant No.\ JA 483/29-1.

\end{document}